\documentclass[twocolumn,showpacs,preprintnumbers,amsmath,amssymb,superscriptaddress,prl]{revtex4}
\usepackage{graphicx}
\usepackage{dcolumn}
\usepackage{bm}
\usepackage{subfigure}
\usepackage{epsfig}

\newcommand{\half}{\mbox{${\textstyle \frac{1}{2}}$}}           
\newcommand{\reactiona}{\mbox{$pp\to ppK^+K^-$}}
\newcommand{\reactionb}{\mbox{$pp\to dK^+\bar{K}^0$}}
\newcommand{\reactionc}{\mbox{$pn\to dK^+K^-$}}
\newcommand{\reactiond}{\mbox{$pd\to p_{\text{sp}} dK^+K^-$}}

\newcommand{\fmn}[2]{\mbox{${\textstyle \frac{#1}{#2}}$}}

\begin{document}
\title{Measurement of the $\bm{pn{\to}dK^+K^-}$ total cross section close to
threshold}

\author{Y.~Maeda}\email[E-mail: ]{ymaeda@rcnp.osaka-u.ac.jp}%
\affiliation{Research Center for Nuclear Physics, Osaka
University, Ibaraki, Osaka 567-0047, Japan}
\author{M.~Hartmann}\email[E-mail: ]{M.Hartmann@fz-juelich.de}%
\affiliation{Institut f\"ur Kernphysik, Forschungszentrum J\"ulich GmbH,
52425 J\"ulich, Germany}
\affiliation{J\"ulich Centre for Hadron Physics, 52425 J\"ulich, Germany}
\author{I.~Keshelashvili}
\affiliation{Department of Physics, University of Basel,
Klingelbergstrasse 82, 4056 Basel, Switzerland}
\author{S.~Barsov}
\affiliation{High Energy Physics Department, Petersburg Nuclear
Physics Institute, 188350 Gatchina, Russia}
\author{M.~B\"uscher}
\affiliation{Institut f\"ur Kernphysik, Forschungszentrum J\"ulich GmbH,
52425 J\"ulich, Germany} \affiliation{J\"ulich Centre for Hadron
Physics, 52425 J\"ulich, Germany}
\author{A.~Dzyuba}
\affiliation{High Energy Physics Department, Petersburg Nuclear
Physics Institute, 188350 Gatchina, Russia}
\author{S.~Dymov}
\affiliation{Laboratory of Nuclear Problems, Joint Institute for
Nuclear Research, 141980 Dubna, Russia}%
\affiliation{Physikalisches Institut II, Universit\"at
Erlangen--N\"urnberg, 91058 Erlangen, Germany} %
\author{V.~Hejny}
\affiliation{Institut f\"ur Kernphysik, Forschungszentrum J\"ulich GmbH,
52425 J\"ulich, Germany} \affiliation{J\"ulich Centre for Hadron
Physics, 52425 J\"ulich, Germany}
\author{A.~Kacharava}
\affiliation{Institut f\"ur Kernphysik, Forschungszentrum J\"ulich GmbH,
52425 J\"ulich, Germany} \affiliation{J\"ulich Centre for Hadron
Physics, 52425 J\"ulich, Germany}
\author{V.~Kleber}
\affiliation{Physikalisches Institut, Universit\"at Bonn, 53115 Bonn,
Germany}
\author{V.~Koptev}
\affiliation{High Energy Physics Department, Petersburg Nuclear
Physics Institute, 188350 Gatchina, Russia}
\author{P.~Kulessa}
\affiliation{H.~Niewodnicza\'{n}ski Institute of Nuclear Physics PAN,
31342 Krak\'{o}w, Poland}
\author{T.~Mersmann}
\affiliation{Institut f\"ur Kernphysik, Universit\"at M\"unster,
48149 M\"unster, Germany}
\author{S.~Mikirtytchiants}
\affiliation{Institut f\"ur Kernphysik, Forschungszentrum J\"ulich GmbH,
52425 J\"ulich, Germany} \affiliation{J\"ulich Centre for Hadron
Physics, 52425 J\"ulich, Germany} \affiliation{High Energy Physics
Department, Petersburg Nuclear Physics Institute, 188350 Gatchina,
Russia}
\author{A.~Mussgiller}
\affiliation{Physikalisches Institut II, Universit\"at
  Erlangen--N\"urnberg, 91058 Erlangen, Germany}
\author{M.~Nekipelov}
\affiliation{Institut f\"ur Kernphysik, Forschungszentrum
  J\"ulich GmbH, 52425 J\"ulich, Germany}
\affiliation{J\"ulich Centre for Hadron Physics, 52425 J\"ulich,
Germany}
\author{H.~Ohm}
\affiliation{Institut f\"ur Kernphysik, Forschungszentrum
  J\"ulich GmbH, 52425 J\"ulich, Germany}
\affiliation{J\"ulich Centre for Hadron Physics, 52425 J\"ulich,
Germany}
\author{K.~Pysz}
\affiliation{H.~Niewodnicza\'{n}ski Institute of Nuclear Physics PAN,
31342 Krak\'{o}w, Poland}
\author{H.J.~Stein}
\affiliation{Institut f\"ur Kernphysik, Forschungszentrum
  J\"ulich GmbH, 52425 J\"ulich, Germany}
\affiliation{J\"ulich Centre for Hadron Physics, 52425 J\"ulich,
Germany}
\author{H.~Str\"oher}
\affiliation{Institut f\"ur Kernphysik, Forschungszentrum
  J\"ulich GmbH, 52425 J\"ulich, Germany}
\affiliation{J\"ulich Centre for Hadron Physics, 52425 J\"ulich, Germany}
\author{Yu.~Valdau}
\affiliation{High Energy Physics Department, Petersburg Nuclear
Physics Institute, 188350 Gatchina, Russia}
\author{C.~Wilkin}
\affiliation{Physics and Astronomy Department, UCL,
Gower Street, London WC1E 6BT, UK}
\author{P.~W\"ustner}
 \affiliation{Zentralinstitut f\"ur Elektronik,
Forschungszentrum J\"ulich GmbH, 52425 J\"ulich, Germany}
\date{\today}

\begin{abstract}
Measurements of the \reactiond\ reaction, where $p_{\text{sp}}$ is a
spectator proton, have been undertaken at the Cooler Synchrotron
COSY--J\"ulich by detecting a fast deuteron in coincidence with a
$K^+K^-$ pair in the ANKE facility. Although the proton beam energy
was fixed, the moving target neutron allowed values of the
non-resonant quasi-free \reactionc\ total cross section to be deduced
up to an excess energy $\epsilon\approx 100~$MeV. Evidence is found
for the effects of $K^-d$ and $K\bar{K}$ final state interactions.
The comparison of these data with those of \reactiona\ and
\reactionb\ shows that all the total cross sections are very similar
in magnitude.

\end{abstract}

\maketitle

We have recently published measurements of the differential and total
cross sections for the \reactiona\ reaction at three energies close
to threshold~\cite{Maeda08}. A major challenge in the analysis was
the separation of the contribution from the production and decay of
the $\phi$ meson from that of the non-$\phi$
component~\cite{Hartmann06}. One of the striking features of the
non-$\phi$ results is the strong attraction between the $K^-$ and
each of the final protons seen in the differential distributions.
This also has a major effect on the energy dependence of the total
cross section, enhancing it at low energies. Although tantalizing,
these results do not, however, resolve the ongoing question as to
whether the interaction is sufficiently strong to allow the $K^-$ to
form a bound state with the two protons~\cite{Agnello,Yamazaki,Gal}.

The isospin dependence of $\phi$ production has been studied through
an investigation of \reactiond~\cite{Maeda06}. By identifying the
final deuteron and kaon pair and measuring their momenta, it was
possible to construct the momentum of the recoil proton
$p_{\text{sp}}$ to show that it was consistent with being a
\emph{spectator}, whose only significant participation in a reaction
is through a change in the kinematics. Interpreting the results in
this way, it was possible to extract values of the quasi-free
\reactionc\ cross section. Moreover, although the experiment was
carried out at one fixed beam energy, the movement of the target
neutron enabled data to be obtained over a wide range of excess
energy $\epsilon=\sqrt{s}-m_{d}-2m_K$ on an event-by-event basis.
Just as in the \reactiona\ case, the shape of the $K^+K^-$ invariant
mass distribution was used to separate the $\phi$ component from the
non-$\phi$ background. The invariant $K^+K^-$ mass spectrum for all
events above the $\phi$ threshold is to be found in
Ref.~\cite{Maeda06} and from this it is already seen that the
non-$\phi$ contribution is a much smaller fraction of the total than
in the \reactiona\ case~\cite{Maeda08}. The prime purpose of this
work is to present the data on the energy dependence of the
non-$\phi$ total cross section up to an excess energy of
$\epsilon\approx 100$~MeV.

Unlike $\phi$ production, there are two different \reactionc\ isospin
channels. The $I=1$ has already been investigated in some detail
through the measurement of \reactionb\ at two beam energies,
corresponding to $\epsilon= 47$ and 105~MeV~\cite{Vera,Alexey2}. The
identical nature of the initial protons, combined with angular
momentum and parity conservation laws, demands that the
$dK^+\bar{K}^0$ final state must contain at least one $p$-wave. At
low energies this will suppress the $I=1$ contribution to \reactionc\
compared to $I=0$ where there is no such constraint. As a
consequence, the energy dependence of the \reactionc\ total cross
section is expected to be more complicated than that of \reactiona.

The investigation was carried using a $2.65~$GeV proton beam incident
on an internal target of the Cooler Synchrotron COSY. The
experimental details and the identification of the $dK^+K^-$
candidates were described for $\phi$ production~\cite{Maeda06} and so
we can here be very brief. The forward-going deuteron was measured in
the ANKE magnetic spectrometer~\cite{ANKE} and both charged kaons
were identified in coincidence on the basis of time-of-flight
criteria. After putting a $\pm 3\sigma$ cut around the missing mass
of the spectator proton, about 4500 $p_{\text{sp}}dK^+K^-$ events
were recorded. The background from misidentified $p\pi^+\pi^-$ events
was estimated to be less than 7\% and effects from this were included
in the systematic uncertainties.

\begin{figure}[ht]
\centering
\includegraphics[clip,width=0.49\columnwidth]{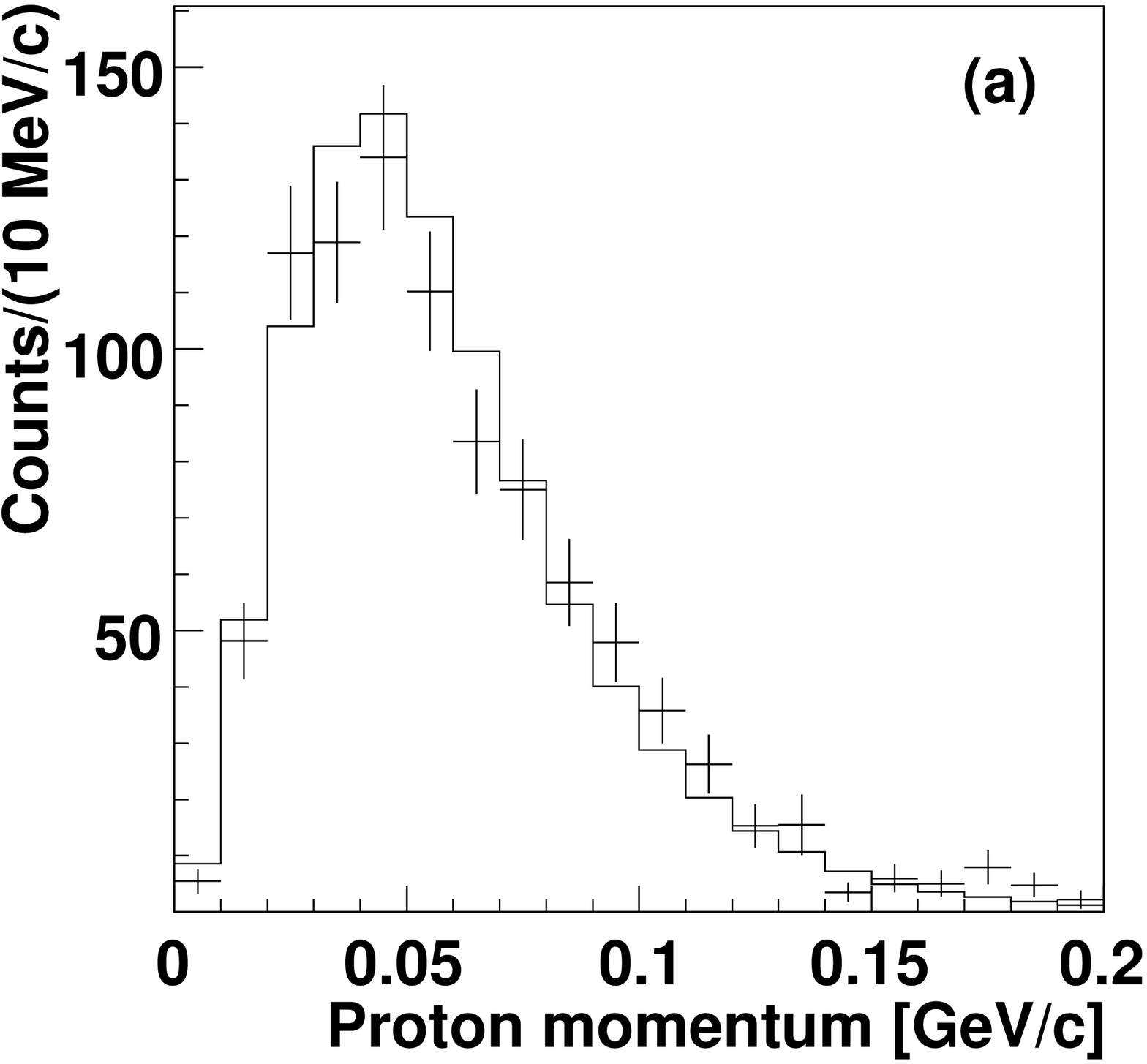}
\includegraphics[clip,width=0.49\columnwidth]{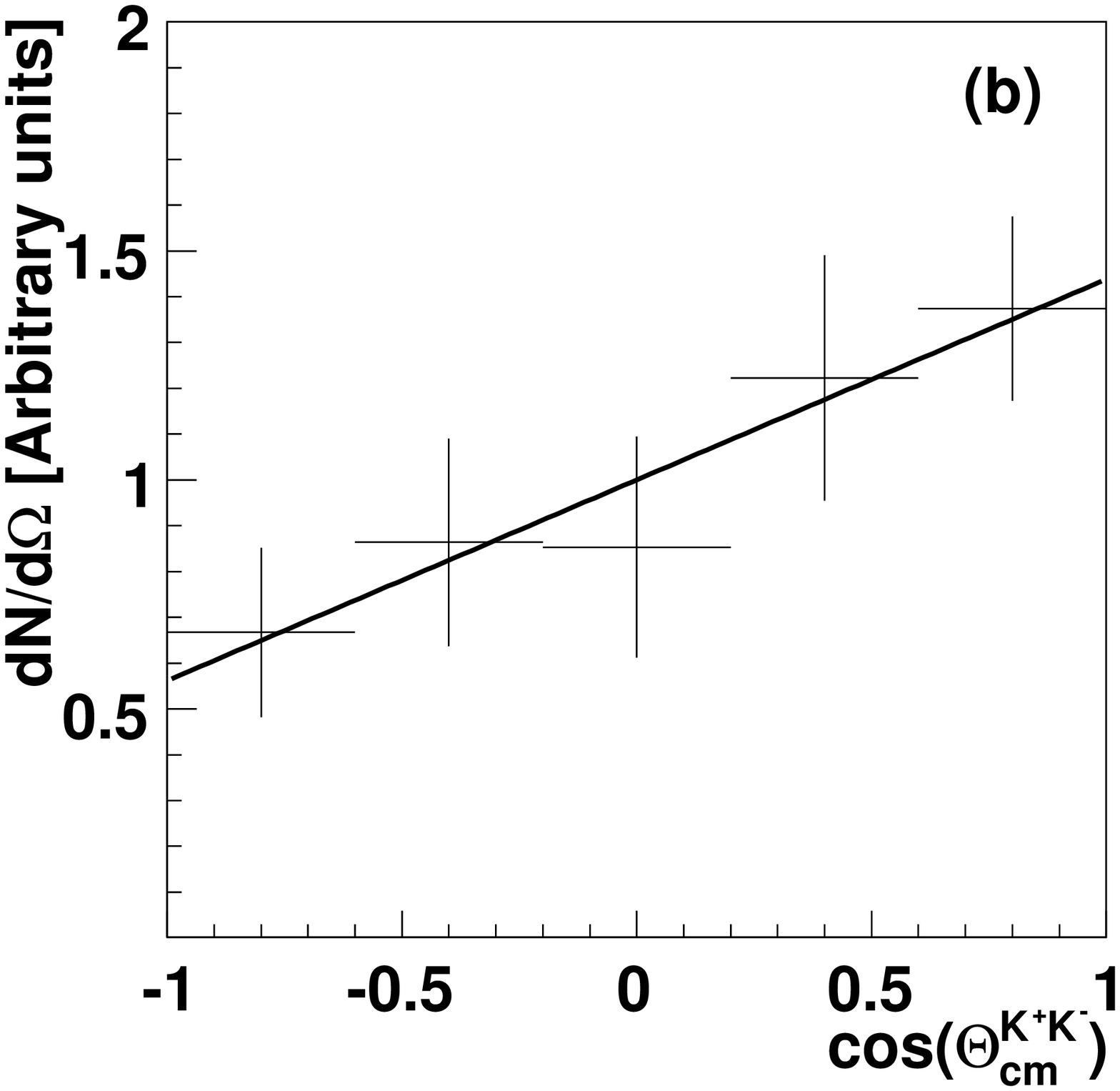}
\caption{(a) Spectator momentum distribution of non-$\phi$ events
compared to simulation using the Bonn potential~\cite{Bonn}. (b)
Acceptance-corrected angular dependence of the polar angle of the
$K^{+}K^{-}$ system relative to the beam axis in the overall c.m.
system at 12$<\epsilon<$32 MeV. The data are well described by
1+0.4~$\cos\theta$ (solid line).} \label{spectate}
\end{figure}

The identification of the residual proton as a spectator is supported
by its momentum distribution shown in Fig.~\ref{spectate}(a),
which follows well the prediction based upon the Bonn wave function~\cite{Bonn}.

The effective target density was determined by measuring the
frequency shift of the stored proton beam as it lost energy due to
its repeated passages through the target~\cite{Zapfe,PRSTAB}.
Combining this with a measurement of the beam current, an integrated
luminosity of $(23\pm 1.4)~\textrm{pb}^{-1}$ was found over the 300
hours of data-taking.

The excess energy determined through the measurement of the momenta
of the deuteron and the two kaons had a standard deviation that was
typically $\sigma_{\epsilon}\approx 2$~MeV, which is small compared
to the 10~MeV bins that were used in the subsequent data analysis. In
order to evaluate the cross section in one of these $\epsilon$
intervals, the geometrical acceptance, resolution, detector
efficiency and kaon decay probability were taken into account in a
Monte Carlo simulation, using the GEANT4 program~\cite{GEANT4}.
The fraction of the total luminosity falling within this interval was
estimated from the deuteron Fermi momentum distribution predicted
using the Bonn potential~\cite{Bonn}.

In the first step of the analysis, the distributions previously
published~\cite{Maeda06} were taken as the basis of the simulation of
the $\phi$--production. For the non--$\phi$ component, three--body
phase space was used. At each excess energy, the four independent
c.m.\ distributions generated were chosen to be the $K^+K^-$
invariant mass as well as three angular distributions. These were
then divided into two groups, depending on the value of the $K^+K^-$
invariant mass, \emph{i.e.}, a $\phi$--rich region where
$1.05<M(K^{+}M^{-})<1.35$~GeV/$c^2$ with the remainder being
designated as the $\phi$--poor region. All distributions were jointly
fitted to the experimental data and the relative contribution of
$\phi$ and non--$\phi$ production evaluated in order to determine the
two acceptances.

\begin{figure}[ht]
\centering
\includegraphics[clip,width=0.49\columnwidth]{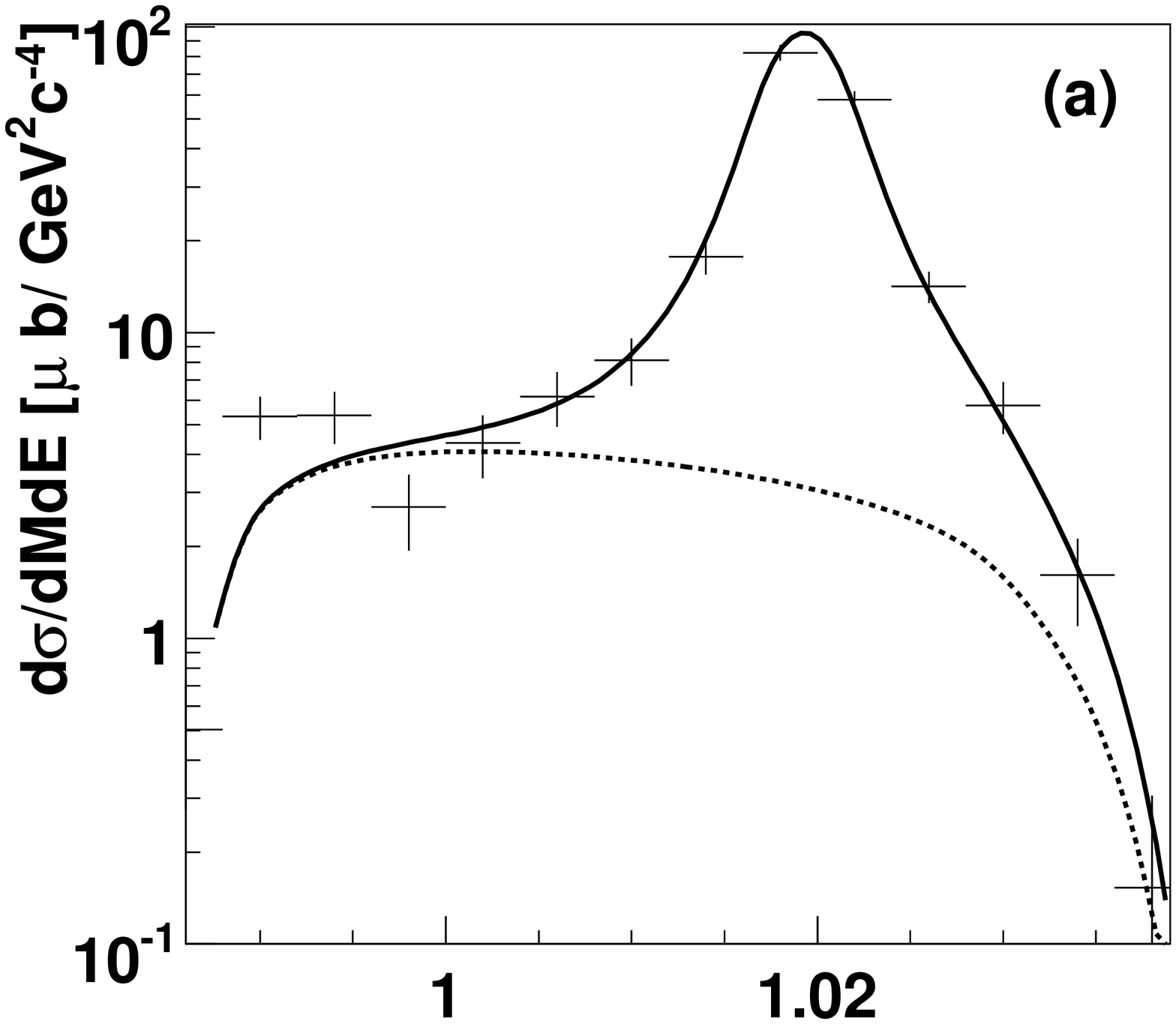}
\includegraphics[clip,width=0.49\columnwidth]{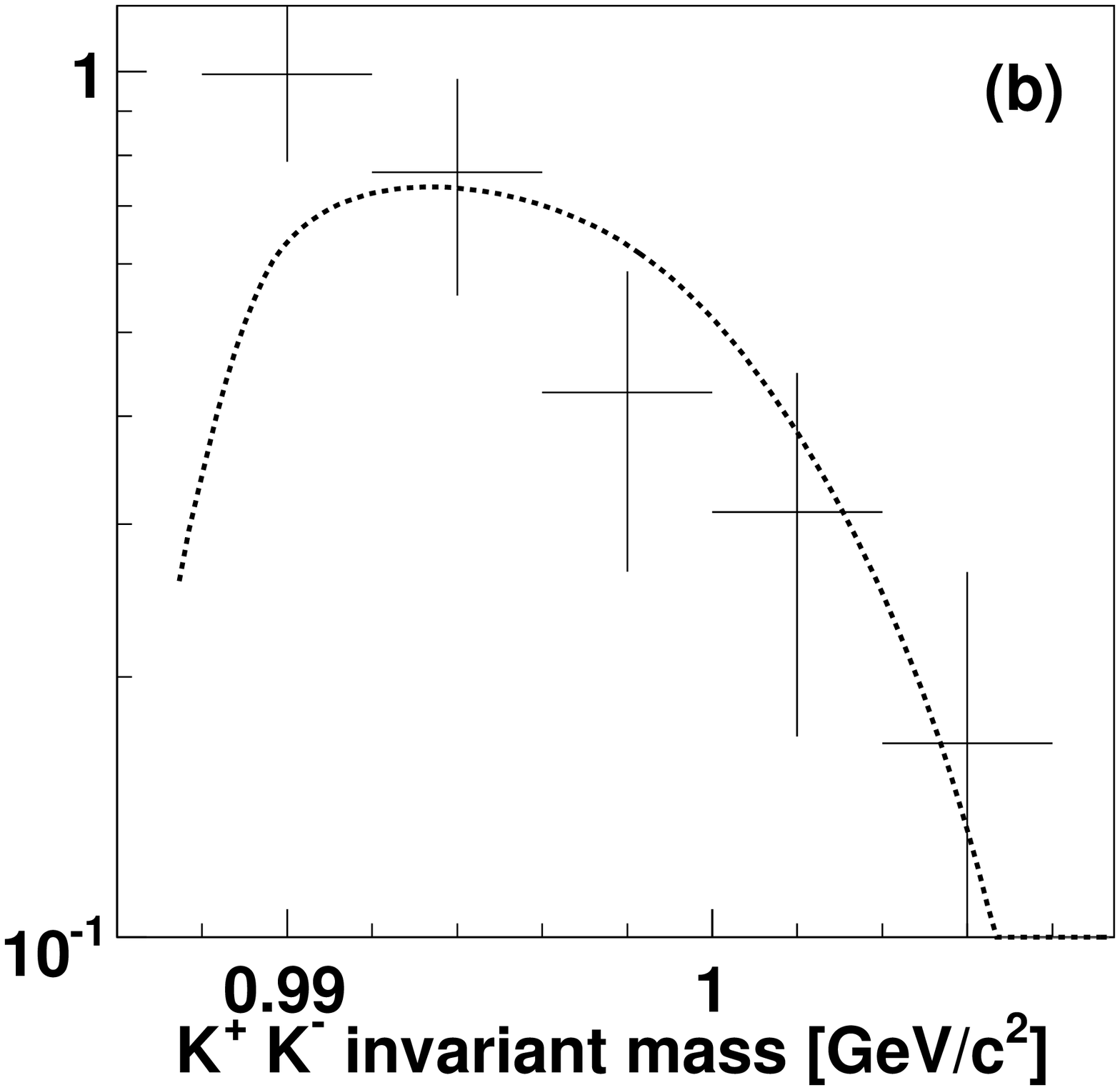}
\caption{$K^+K^-$ invariant mass distributions of the \reactionc\
reaction for event samples below and above the $\phi$ threshold. The
data are shown for excess energy bins (a) $42<\epsilon<52$~MeV, and
(b) $12<\epsilon<22$~MeV. The non--$\phi$ simulation (dashed curve)
includes the effect of the $K^-d$ \emph{fsi}. The solid curve
includes also a contribution from $\phi$ production, where
applicable.}\label{fig1}
\end{figure}

\begin{figure}[htb]
\begin{center}
\epsfig{file=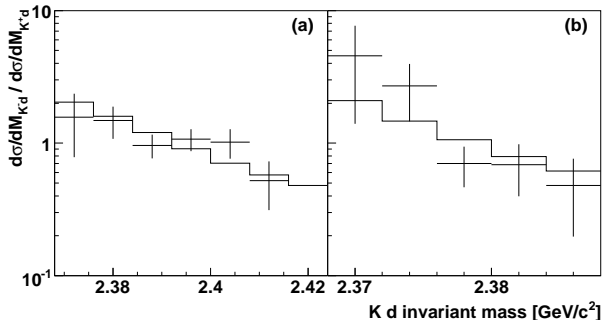,width=1.05\columnwidth}
\end{center}
\vspace{-7mm} \caption{Ratios of the $K^{\pm}d$ invariant mass
distributions for the same two ranges of excess energy as in
Fig.~\ref{fig1}. (a): $42<\epsilon<52$~MeV, (b): $12<\epsilon<22$~MeV.
 The histograms are the simulations in the
scattering length approximation with $a = (-1.0+i1.2)$~fm.}
\label{fig2}
\end{figure}

In the $\phi$--poor region, the polar angle of the $K^+K^-$ system
relative to the beam axis in the overall c.m.\ system shows a forward
peak for $\epsilon<$ 52~MeV and a typical acceptance-corrected
distribution is shown in Fig.\ref{spectate}(b).
The forward/backward asymmetry is evidence for some $I=0/I=1$ interference.
The angular distribution for the $\phi$--rich region is fairly symmetrical
because of the $I=0$ dominance and the lack of interference between
the $\phi$ and the background.
 Fitting the shape with $1+\alpha\cos\theta$,
one finds $\alpha=0.4\pm0.1$ at low energies but $\alpha$ consistent
with zero at higher $\epsilon$.
Taking $\alpha$ to have the polynomial energy dependence on
$\epsilon$, its inclusion increases the total acceptance for non--$\phi$
production up to 8.5\%.
 In addition, as discussed below, the
$K^{\pm}d$ invariant masses deviate from phase space due to the
strong final state interaction between $K^-$ and deuteron
(Fig.~\ref{fig2}). This was included through a $K^{-}d$ enhancement
factor based on a scattering length approximation~\cite{Alexey2}. The
deviations were taken into account iteratively in the simulations in
order to converge on acceptance--corrected distributions.

Two typical $K^+K^-$ mass spectra from above and below the $\phi$
threshold are shown in Fig.~\ref{fig1} after making acceptance and
other corrections. Also illustrated there are fits to the $\phi$ and
non-$\phi$ contributions to the cross section where, in the latter
case, a distorted three-body phase space has been assumed for the
$dK^+K^-$ final state. In general the $\phi$ contribution is well
described but the same cannot be said for the non-$\phi$
distribution. The difficulties here arise principally from the
unknown fraction of $p$-waves that come from the $I=1$ cross section
and the influence of the $K^-d$ final state interaction. In addition,
for invariant masses below about 995~MeV/$c^2$, the data in
Fig.~\ref{fig1}(a) lie well above the simulation. This feature, which
is also seen quite clearly in the \reactiona\
data~\cite{Maeda08,DISTO}, is evidence for a final state interaction
in the $K\bar{K}$ subsystem. In the region of the $K^0\bar{K}^0$
threshold, the $K^+K^-\rightleftharpoons K^0\bar{K}^0$ coupling can
lead to a cusp effect~\cite{Alexey3}.

Of much more importance for the acceptance estimation is the
distortion in the $K^-d$ subsystem. Figure \ref{fig2} shows the ratio
of the differential cross sections
\begin{equation}
R_{Kd}=\frac{d\sigma/dM(K^-d)}{d\sigma/dM(K^+d)}
\end{equation}
for the non--$\phi$ region at the same excess energies as those shown
in Fig.~\ref{fig1}. Here $M(K^{\pm}d)$ is the invariant mass of the
$K^{\pm}d$ subsystem. Experimental distributions of $R_{Kd}$ at both
energies show a very strong preference for low values of $M(Kd)$,
which arises from the interaction between $K^-$ and deuteron. A
$K^{-}d$ final state interaction factor is introduced into the
three--body phase space simulation of the non--$\phi$ contribution by
using the scattering length approximation $1/(1-iqa)$, where $q$ is
the $K^-d$ relative momentum. The (complex) scattering length is
believed to be of the order of $a \approx
(-1.0+i1.2)$\,fm~\cite{Rusetsky}, which would correspond a bound or
virtual state with a binding energy of $\epsilon_0\approx 20$\,MeV.
The ANKE $pp\to dK^+\bar{K}^0$ data seem to be insensitive to the
phase of $a$ but they are best fit with $|a|\approx
1.5$~fm~\cite{Alexey2}. After taking $a=(-1.0+i1.2)$\,fm, the
individual $d\sigma/dM(K^+d)$ and $d\sigma/dM(K^-d)$ distributions
are well described, as is the ratio $R_{Kd}$, which is shown for the
two excess energy intervals by the histograms in Fig.~\ref{fig2}.

In view of the low statistics and the consequent fluctuations, the
non-$\phi$ total cross section in a $\epsilon$ bin was evaluated in
two different ways, (a) by subtracting the fit to the $\phi$
component in Fig.~\ref{fig1} and summing the remainder, and (b) by
taking the direct fit to the non-$\phi$ part of Fig.~\ref{fig1}. The
average of these two values is given as the total cross section in
Table~\ref{table1}. The difference is a major contributor to the
systematic uncertainties given there.

\begin{table}[hbt]
\caption{\label{table1}%
Total cross section for the non-$\phi$ component of the
\reactionc\ reaction as a function of the excess energy $\epsilon$
with respect to the $dK^+K^-$ threshold. The first error on the
cross section is statistical and the second systematic whereas
that on the energy is the bin half-width. The overall $\approx\pm
6\%$ uncertainty in the luminosity has not been compounded with
the other errors.}
\begin{ruledtabular}
\begin{tabular}{cl}
$\varepsilon$&\hspace{3mm}
$\sigma_{\text{non-}\phi}$(tot)\\
(MeV)&\hspace{9mm}(nb)\\
\hline
 $17.1\pm5.0$&  $\phantom{1}1.8\pm0.3\pm0.1$\\
 $27.1\pm5.0$&  $\phantom{1}5.9\pm1.0\pm0.7$\\
 $37.1\pm5.0$&  $12.4\pm2.5\pm1.8$\\
 $47.1\pm5.0$&  $16.2\pm3.6\pm1.9$\\
 $57.1\pm5.0$&  $27.6\pm4.8\pm3.2$\\
 $67.1\pm5.0$&  $34.9\pm6.7\pm3.4$\\
 $77.1\pm5.0$&  $38.6\pm10.0\pm5.0$\\
 $87.1\pm5.0$&  $50.2\pm14.3\pm7.7$\\
$102.1\pm10.0$& $69.5\pm17.5\pm10.7$\\
\end{tabular}
\end{ruledtabular}
\end{table}

\begin{figure}[ht]
\centering
\includegraphics[clip,width=0.8\columnwidth]{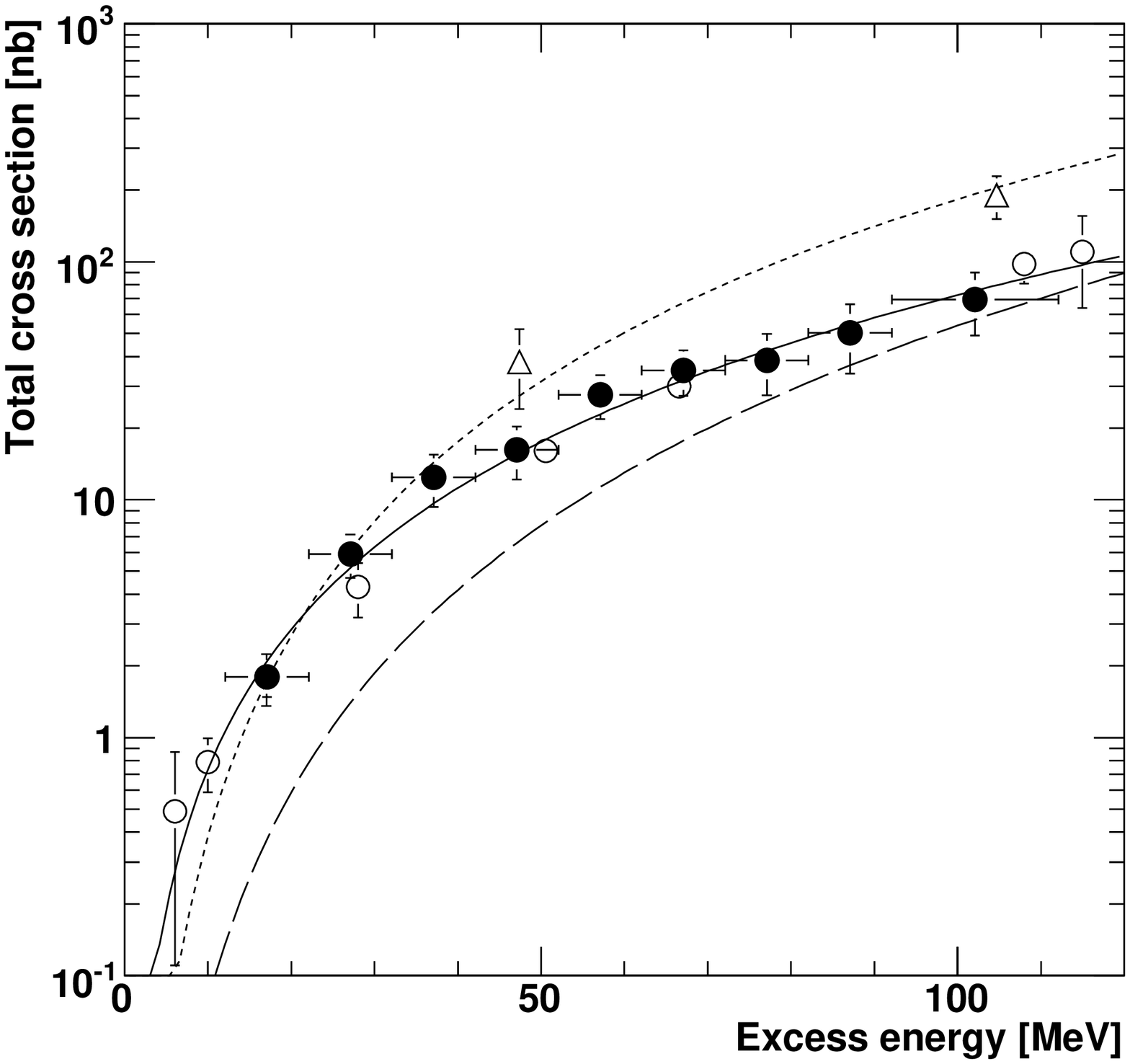}
\caption{Total cross section for non-$\phi$ $K\bar{K}$ production in
nucleon-nucleon collisions near threshold. The closed circles denote
\reactionc\ data (this work) and \reactionb\ (open triangles
\cite{Vera}), whereas the open circles show the results for
\reactiona\ from Refs.~\cite{Maeda08,DISTO,COSY11}. The dotted curve
is the best fit of Eq.~(\ref{s1}) to the \reactionb\ data whereas the
solid curve includes also the isospin-zero contribution of
Eq.~(\ref{s0}) so as to describe the energy dependence of the
\reactionc\ total cross section. The dashed curve represents the
subsequent prediction of Eq.~(\ref{ratio}) for the $pn\to
\{pn\}_{I=0}K^+K^-$ total cross section. }\label{totalxs}
\end{figure}

The values of the \reactionc\ total cross section presented in
Fig.~\ref{totalxs} shows a smooth behavior on a logarithmic scale.
Also shown are results for the \reactiona~\cite{Maeda08,DISTO,COSY11}
and \reactionb~\cite{Vera}.

The isospin dependence of kaon pair production can be deduced from
\begin{eqnarray}
\nonumber \sigma(\reactionb)&=&\sigma_1\,,\\
\sigma(\reactionc)&=&\fmn{1}{4}(\sigma_1+\sigma_0)\,. \label{isospin}
\end{eqnarray}
Interpolating our results to the energies where the \reactionb\ has
been studied~\cite{Vera}, we find isospin ratios of
$\sigma_0/\sigma_1 = 0.9\pm 0.9$ at 47~MeV and $0.5\pm0.5$ at
105~MeV. The large error bars arise from the subtraction implicit in
Eq.~(\ref{isospin}) and it is hard to draw firm conclusions except
that $\sigma_0$ cannot be much larger than $\sigma_1$. There might be
a tendency for $\sigma_1$ to become relatively more important as the
energy is raised. This is what one would expect from the
requirement of having a $p$-wave in the \reactionb\ final
state~\cite{Vera,Alexey2}.

A major influence on the energy dependence of the total cross section
arises from the $K^-\,d$ final state interaction which gives the
distortion shown in Fig.~\ref{fig2}. Although the closed form
description of Ref.~\cite{FW1} is only strictly valid for a real
scattering length, to a good approximation the energy dependence of
the two cross sections should be given by
\begin{eqnarray}
\nonumber
\label{s0}\sigma_0 &=& A_0\,\epsilon^2/D\,,\\
\label{s1}\sigma_1&=&
A_1\,\epsilon^3\left(D+\half\epsilon/\epsilon_0\right)/D^2\,,\\
D&=&\left(1+\sqrt{1+\epsilon/\epsilon_0}\right)^{\!2},\nonumber
\end{eqnarray}
where $\epsilon_0\approx 20$~MeV.

The choice of the $I=0$ and $I=1$ coefficients $A_0=127$~pb/MeV$^2$
and $A_1=1.8$~pb/MeV$^3$ leads to the fits to the \reactionb\ and
\reactionc\ total cross sections shown in Fig.~\ref{totalxs}. The
general behavior is reproduced much better than it would be if the
$K^-d$ final state interaction were neglected.

Another important feature of Fig.~\ref{totalxs} is that the
\reactiona\ and \reactionc\ total cross sections are similar in
magnitude. However some allowance has to be made for the four-body
nature of the $ppK^+K^-$ phase space. An estimate of this effect
can be obtained in a simple final state interaction
model~\cite{WTF}. This predicts that
\begin{eqnarray}
\nonumber \lefteqn{\left.\sigma(pn\to
\{pn\}_{I=0}K^+K^-)\right/\sigma(pn\to dK^+K^-) \approx}\\
&&\hspace{-4mm}\frac{2}{\pi\sqrt{x}}\left[\frac{5}{6}+\frac{4x}{15}
+\frac{1}{2x}-2\sqrt{x}\left(\frac{1+x}{2x}\right)^{\!2}
\textrm{arctan}\sqrt{x}\right], \label{ratio}
\end{eqnarray}
where $x=\epsilon/B$, with $B$ denoting the deuteron binding
energy. The result of multiplying this ratio by the fit to the
\reactionc\ total cross section is shown in Fig.~\ref{totalxs}.
Using this simple estimate we see that
\begin{equation}
\sigma(\reactiona)/\sigma(pn\to \{pn\}_{I=0}K^+K^-) \approx 1.5\,.
\end{equation}

It is clear from the results presented here that, after correcting
for the different phase spaces, the total cross sections for the
\reactiona, \reactionc, and \reactionb\ reactions are very similar in
magnitude despite the necessity for $p$-waves in the last case. It
would be highly desirable to have a common theoretical model to
describe all three channels.
\\

\begin{acknowledgments}
We are grateful for the support offered by the COSY machine crew and
other members of the ANKE Collaboration. This work has been partially
financed by the BMBF, DFG, Russian Academy of Sciences, HGF--VIQCD,
and COSY FFE.
\end{acknowledgments}
%
%


\begin{thebibliography}{99}
%
\bibitem{Maeda08} Y.~Maeda \emph{et al.}, Phys.\ Rev.\ C \textbf{77},
015204 (2008).
%
\bibitem{Hartmann06} M.~Hartmann \emph{et al}., Phys.\ Rev.\ Lett.\ \textbf{96}, 242301 (2006).
%
\bibitem{Agnello} M.~Agnello \emph{et al.}, Phys.\ Rev.\ Lett.\
\textbf{94}, 212303 (2005).
%
\bibitem{Yamazaki} T.~Yamazaki and Y.~Akaishi, Phys.\ Rev.\ C \textbf{76}, 045201 (2007).
%
\bibitem{Gal} N.V.~Shevchenko, A.~Gal, J.~Mare\v{s}, J.~R\'{e}vai,
Phys.\ Rev.\ C \textbf{76}, 044004 (2007).
%
\bibitem{Maeda06} Y.~Maeda \emph{et al}., Phys.\ Rev.\ Lett.\ \textbf{97}, 142301 (2006).
%
\bibitem{Vera} V.~Kleber \emph{et al.}, Phys.\ Rev.\ Lett.\ \textbf{91},
172304 (2003); A.~Dzyuba \emph{et al.}, Eur.\ Phys.\ J.\ A
\textbf{29}, 245 (2006)
%
\bibitem{Alexey2} A.~Dzyuba \emph{et al.}, Eur.\ Phys.\ J.\ A \textbf{38}, 1 (2008).
%
\bibitem{ANKE}
S.~Barsov \emph{et al.}, Nucl.\ Instrum.\ Methods Phys.\ Res.\ A
\textbf{462}, 364 (2001).
%
\bibitem{Bonn}
R.~Machleidt, K.~Holinde, and Ch.~Elster, Phys.\ Rep.\ \textbf{149},
1 (1987).
%
\bibitem{Zapfe} K.~Zapfe, Nucl.\ Instrum.\ Methods Phys.\ Res.\ A \textbf{368}, 293 (1996).
%
\bibitem{PRSTAB} H.J.~Stein \emph{et al.}, Phys.\ Rev.\ ST Accel.\ Beams, \textbf{11}, 052801 (2008).
%
\bibitem{GEANT4} S. Agostinelli \emph{et al.}, Nucl.\ Instrum.\ Meth.\ A
\textbf{506}, 250 (2003); http://geant4.web.cern.ch/geant4/
%
\bibitem{DISTO} F.~Balestra \emph{et al.}, Phys.\ Rev.\ C \textbf{63}, 024004 (2001).
%
\bibitem{Alexey3} A.~Dzyuba \emph{et al.}, Phys.\ Lett.\ B \textbf{668}, 315
(2008).
%
\bibitem{Rusetsky} U.-G.~Mei{\ss}ner, U.~Raha, A.~Rusetsky, Eur.\ Phys.\ J.\ C
\textbf{47}, 473 (2006).
%
\bibitem{COSY11} M.~Wolke, PhD thesis, University of M\"unster
(1997); C.~Quentmeier \emph{et al.}, Phys.\ Lett.\ B \textbf{515},
276 (2001); P.~Winter \emph{et al.}, Phys.\ Lett.\ B \textbf{635},
23 (2006).
%
%
\bibitem{FW1} G.~F\"aldt and C.~Wilkin, Phys.\ Lett.\ B
\textbf{382}, 209 (1996).
%
\bibitem{WTF} C.~Wilkin, U.~Tengblad, and G.~F\"{a}ldt,
    Acta Physica Slovaca \textbf{56}, 205 (2006).
%
\end{thebibliography}
\end{document}